\RequirePackage{fix-cm}
\documentclass{svjour3}                     
\smartqed  

\usepackage{graphicx}
\usepackage[numbers]{natbib}
\usepackage{dcolumn}
\usepackage{booktabs}

\begin{document}

\title{The Role of Substrate Roughness in Superfluid Film Flow Velocity
}


\author{Jun Usami         \and
        Nobuyuki Kato, Tomohiro Matsui, Hiroshi Fukuyama 
}


\institute{Department of Physics, The University of Tokyo \at
              7-3-1, Hongo, Bunkyo-ku, Tokyo 133-0033, Japan \\
           J. Usami \at
              \email{jusami@kelvin.phys.s.u-tokyo.ac.jp}           
          \and
           H. Fukuyama \at
              \email{hiroshi@phys.s.u-tokyo.ac.jp}
}

\date{Received: date / Accepted: date}

\maketitle

\begin{abstract}
It is known that the apparent film flow rate $j_0$ of superfluid $^4$He increases significantly when the container wall is contaminated by a thin layer of solid air. 
However, its microscopic mechanism has not yet been clarified enough. 
We have measured $j_0$ under largely different conditions for the container wall in terms of surface area (0.77--6.15~m$^2$) and surface morphology using sintered silver fine powders (particle size: $0.10$ ($\mathrm{\mu}$m) and porous glass (pore size: 0.5, 1 $\mathrm{\mu}$m). 
We could increase $j_0$ by more than two orders of magnitude compared to non-treated smooth glass walls, where liquid helium flows down from the bottom of container as a continuous stream rather than discrete drips. 
By modeling the surface morphology, we estimated the effective perimeter of container $L_{\mathrm{eff}}$ and calculated the flow rate $j~(= j_0L_0/L_{\mathrm{eff}})$, where $L_0$ is the apparent perimeter without considering the microscopic surface structures.
The resultant $j$ values for the various containers are constant each other within a factor of four, suggesting that the enhancement of $L_{\mathrm{eff}}$ plays a major role to change $j_0$ to such a huge extent and that the superfluid critical velocity, $v_{\mathrm{c}}$, does not change appreciably.
The measured temperature dependence of $j$ revealed that $v_{\mathrm{c}}$ values in our experiments are determined by the vortex depinning model of Schwarz (Phys. Rev. B {\bf31}, 5782 (1986)) with several nm size pinning sites. 


\end{abstract}

\section{Introduction}
\label{intro}
Film flow is one of the extraordinary phenomena of superfluidity, and is known as a popular demonstration experiment. 
Below the Lambda transition temperature ($T_\mathrm{\lambda} = 2.1768$~K), liquid $^4$He flows out from a container with an open top through a helium thin film adsorbed on the wall \cite{Daunt1938}. 
The film thickness is typically of the order of 10~nm \cite{Everitt1964}.
It is known that the apparent flow rate $j_0$, the superfluid mass flow per unit time divided by the macroscopic perimeter of the container $L_0$, increases dramatically by contaminating the container wall by a thin layer of solid air~\cite{Bowers1950}. 
Similar but much smaller increases were observed on unpolished metal surfaces such as stainless steel, while flow rates on highly polished ones are nearly the same as that on clean glass~\cite{Chandrasekhar1952}. 
This indicates that $j_0$ is insensitive to the wall material but dependent on the microscopic structure of the surface. 

Smith and Boorse \cite{Smith1955} examined the roughness dependence of $j_0$ for various metal surfaces. Their results are consistent with the model in which the $j_0$ enhancement is explained only by the geometrical effect, i.e., the effectively extended perimeter due to the roughness ($\it{effective~perimeter~model}$)~\cite{Bowers1950,Mendelssohn1950}.
However, one can also imagine that $j_0$ in the air contaminated container may also be enhanced by the dramatic increase of the superfluid critical velocity $v_{\mathrm{c}}$ which could happen for some reason. 
To test this possibility, it is crucial to estimate the former effect quantitatively.
However, it is generally difficult to control and evaluate the surface structure of the thin layer of solid air. 

In this paper, we present results of film flow rate measurements of superfluid $^4$He using two different types of glass containers, i.e., glass containers covered by layers of sintered silver powders (Type-1) and those with porous surfaces (Type-2), and compared them with published results of bare glass containers~\cite{Mendelssohn1950,Smith1955a}.
For the surface decorated containers, we observed dramatic increases of $j_0$ at $T =$2.04~K by more than two orders of magnitude compared to that for the bare glass containers.
Then, we evaluated effectively enhanced perimeters $L_\mathrm{eff}$ for them by modeling the surface morphology based on scanning electron microscope (SEM) images and surface area measurements. 
Flow rates $j$ corrected by $L_\mathrm{eff}$, i.e., $j = j_0L_0/L_{\mathrm{eff}}$, for various containers are consistent with each other within a factor of four.
This means that $v_{\mathrm{c}}$ is more or less unchanged regardless of such largely different surface roughness and that the geometrical effect plays a major role on the huge enhancement of $j_0$ in our experiments and most likely in air contaminated containers as well.
The measured temperature dependence of $j$ suggests that $v_{\mathrm{c}}$ is determined by 
the vortex depinning mechanism~\cite{Schwarz1985} with pinning sites smaller than 10~nm.

\section{Experimental methods}

The Type-1 containers are made of Pyrex glass with dimensions; 28/32~mm in inner/outer diameter and 15~mm in inner height.
The rim of an open top of the container (bucket-type) was made as smooth as possible.
The whole surface of the container except 10~mm wide vertical slits to observe inner liquid level was covered by a layer of sintered Ag powder of 0.10~$\mathrm{\mu}$m particle diameter (C-8 Ag powder manufactured by Tokuriki Honten, Co., Ltd.). 
To fix the Ag powder on the glass surface, they were mixed in a paste containing Ag flakes of $\sim5~$$\mathrm{\mu}$m size and painted on the surfaces by a brush.
The weight ratio of Ag powder to paste was 1 (Ag-2) or 5 (others).
Then the container was heated in air to sinter the Ag powder at different temperatures of 120, 150 or 180~$^\circ$C for 15 or 60~min.
We prepared four containers of this type (Ag-1$\sim$Ag-4) with different sintering conditions as shown in Table 1.

\newcolumntype{d}[1]{D{.}{\cdot}{#1}}
\newcolumntype{.}{D{.}{.}{-1}}
\newcolumntype{,}{D{,}{,}{-1}}

\begin{table}[ht]
 \caption{Specifications of various containers. $p$ and $q$ are parameters used in eq.(\ref{eqLeff3}) which models the morphology of porous glass. They correspond to the apparent pore and channel diameters, respectively, in the SEM image (see the main text for more details). In the last line, measured apparent flow rates $j_0$ at $T=2.05$~K are also shown where the data of the bare glass container is from Ref.\cite{Mendelssohn1950,Smith1955a}.}
  \begin{tabular}{llc.....} 
   \toprule\toprule
    & \multicolumn{1}{c}{bare} &\multicolumn{4}{c}{Type-1} &\multicolumn{2}{c}{Type-2}\\
\cmidrule(lr){2-2}\cmidrule(lr){3-6}\cmidrule(lr){7-8}
    &  \multicolumn{1}{c}{Glass} & \multicolumn{1}{c}{Ag-1} & \multicolumn{1}{c}{Ag-2} & \multicolumn{1}{c}{Ag-3} & \multicolumn{1}{c}{Ag-4} & \multicolumn{1}{c}{PG-1} & \multicolumn{1}{c}{PG-2} \\
   \midrule
    ~~$T_{\mathrm{sint}}$ ($^\circ$C) &\multicolumn{1}{c}{---}& 180 & 180 & 150 & 120 &\multicolumn{1}{c}{---}&\multicolumn{1}{c}{---}\\[3pt]
   ~~$t_{\mathrm{sint}}$ (min) &\multicolumn{1}{c}{---}& 60 &60&15&15&\multicolumn{1}{c}{---}&\multicolumn{1}{c}{---}\\
      ~~$p$ ($\mathrm{\mu}$m) &\multicolumn{1}{c}{---}&---&\multicolumn{1}{c}{---}&\multicolumn{1}{c}{---}&\multicolumn{1}{c}{---}& 0.5 & 1\\
   ~~$q$ ($\mathrm{\mu}$m) &\multicolumn{1}{c}{---}&---&\multicolumn{1}{c}{---}&\multicolumn{1}{c}{---}&\multicolumn{1}{c}{---}& 0.3 & 0.5\\[3pt]
      \begin{tabular}{l} $j_0$\ $\times$10$^5$\\~(kg/m/s) \end{tabular} & 0.032 & 4.7, 13, 19 &15 & 1.6 & 3.5 & 3.5 & 2.2 \\
   ~~$S$ (m$^2$) & 0.0044 & 0.79, 1.70, 2.52 & 2.81 & 0.77 & 1.33 & 6.2 & 2.0\\
     \bottomrule\bottomrule
        \end{tabular}
        \label{table}
        \end{table}

\begin{figure}[b]
   \begin{tabular}{c}
 \begin{minipage}{0.33\textwidth}
\begin{flushleft}
 \includegraphics[width=36mm]{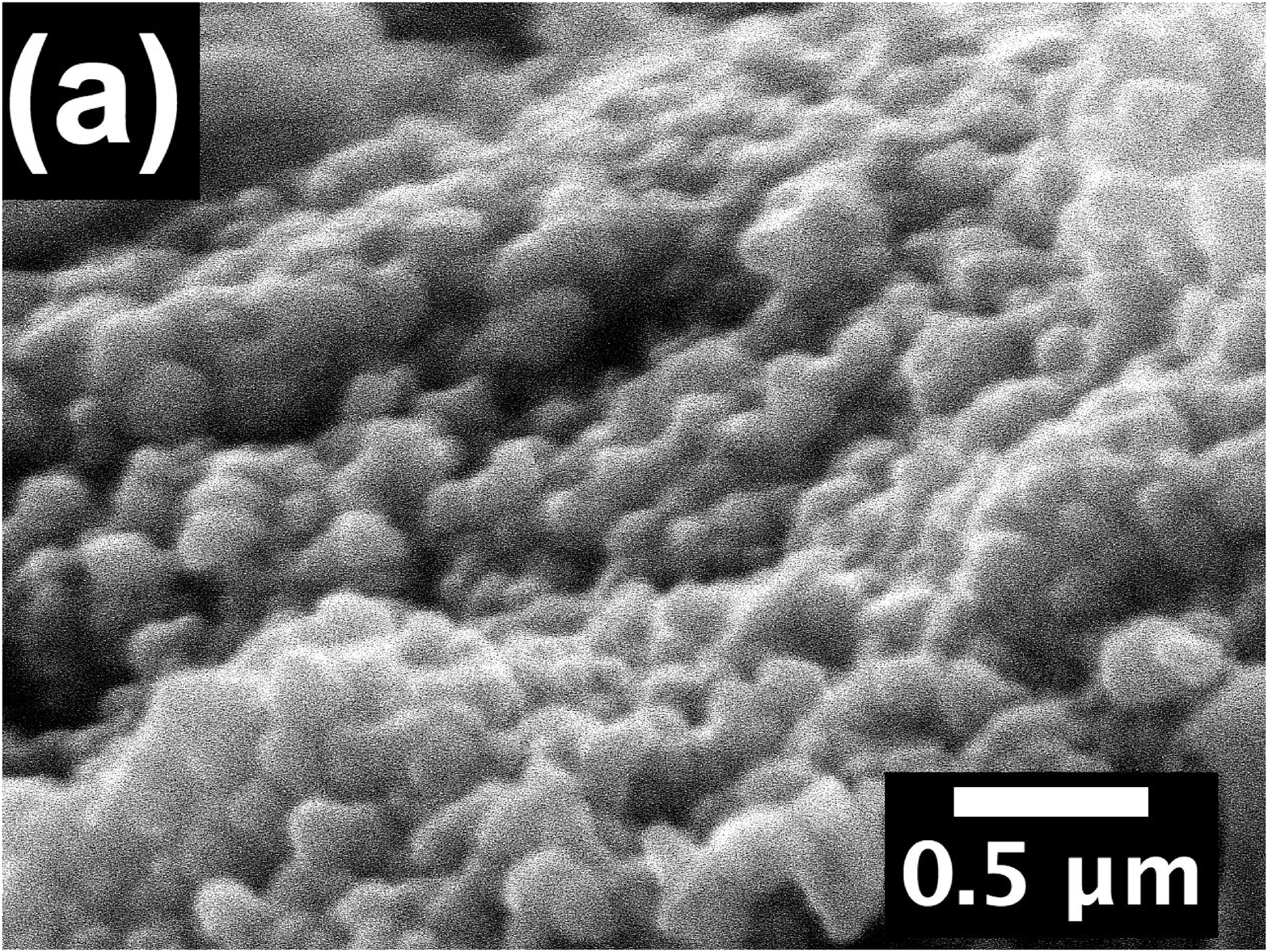}
 \end{flushleft}
 \end{minipage}
 \begin{minipage}{0.33\textwidth}
\begin{flushleft}
 \includegraphics[width=36mm]{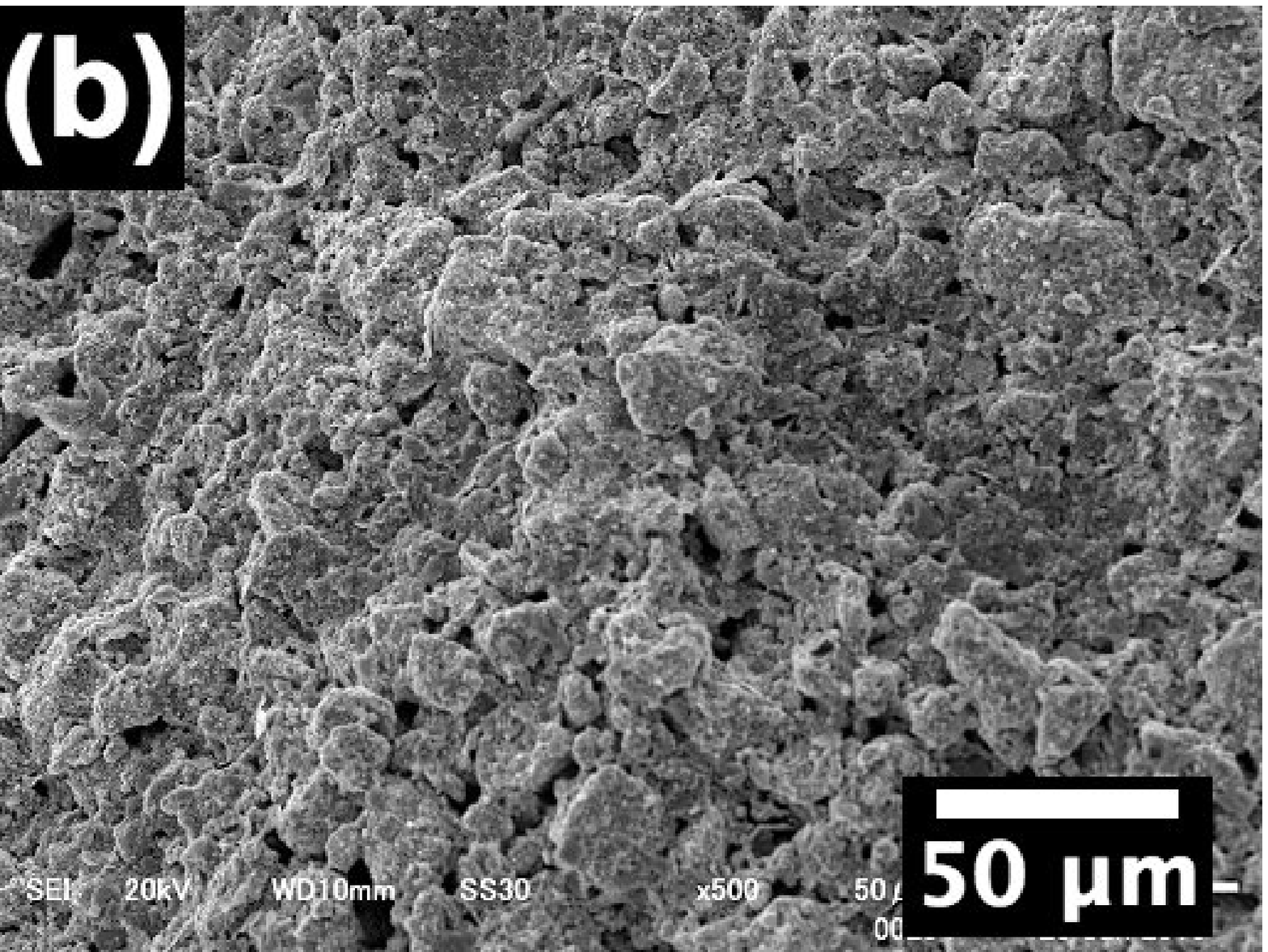}
 \end{flushleft}
 \end{minipage}
 \begin{minipage}{0.33\textwidth}
\begin{flushleft}
 \includegraphics[width=36mm]{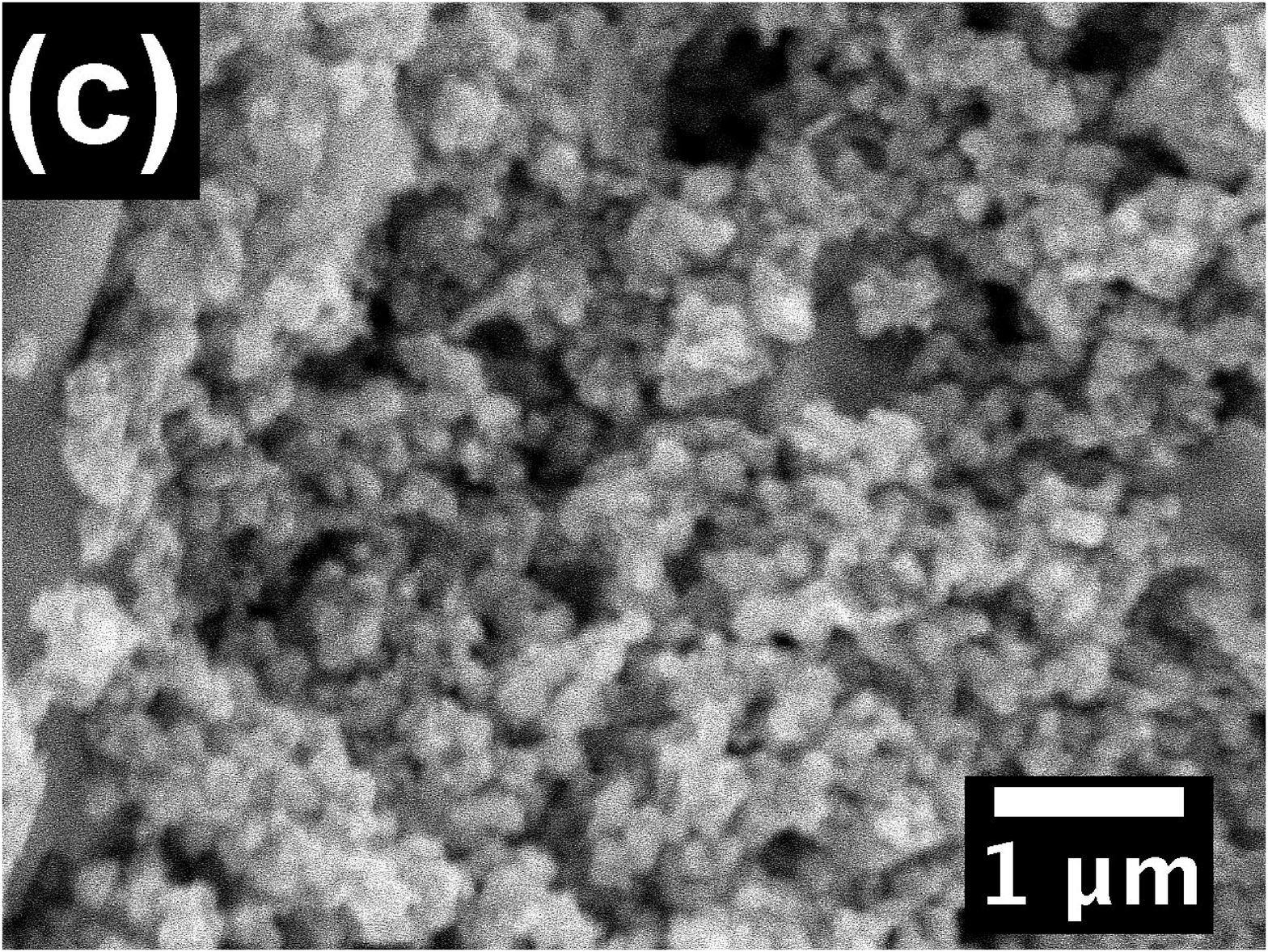}
 \end{flushleft}
 \end{minipage}
  \end{tabular}
   \caption{SEM images of surfaces of (a)(b) Ag-1 and (c) Ag-4 containers of Type-1.}
 \label{photo-Ag}
    \end{figure}

SEM images of surfaces of Ag-1 and Ag-4 are shown in Fig.~\ref{photo-Ag}.
Even at the lowest sintering temperature and for the short sintering time ($T_{\mathrm{sint}} =$~120~$^\circ$C, $t_{\mathrm{sint}} =$~15~min), individual Ag particles and their grains are well connected with each other as seen in Fig.~\ref{photo-Ag}(c).
At the highest sintering temperature and for the long sintering time (180~$^\circ$C, 60~min), as shown in Fig.~\ref{photo-Ag}(a), the inter-particle neck becomes thicker, but the original particle size is still kept not clustering too much.
More global connections are seen in Fig.~\ref{photo-Ag}(b).
In both cases, Ag clusters of a few tens~$\mathrm{\mu}$m size are well connected with each other.

  \begin{figure}[t]
   \begin{tabular}{c}
 \begin{minipage}{0.33\textwidth}
\begin{flushleft}
 \includegraphics[width=36mm]{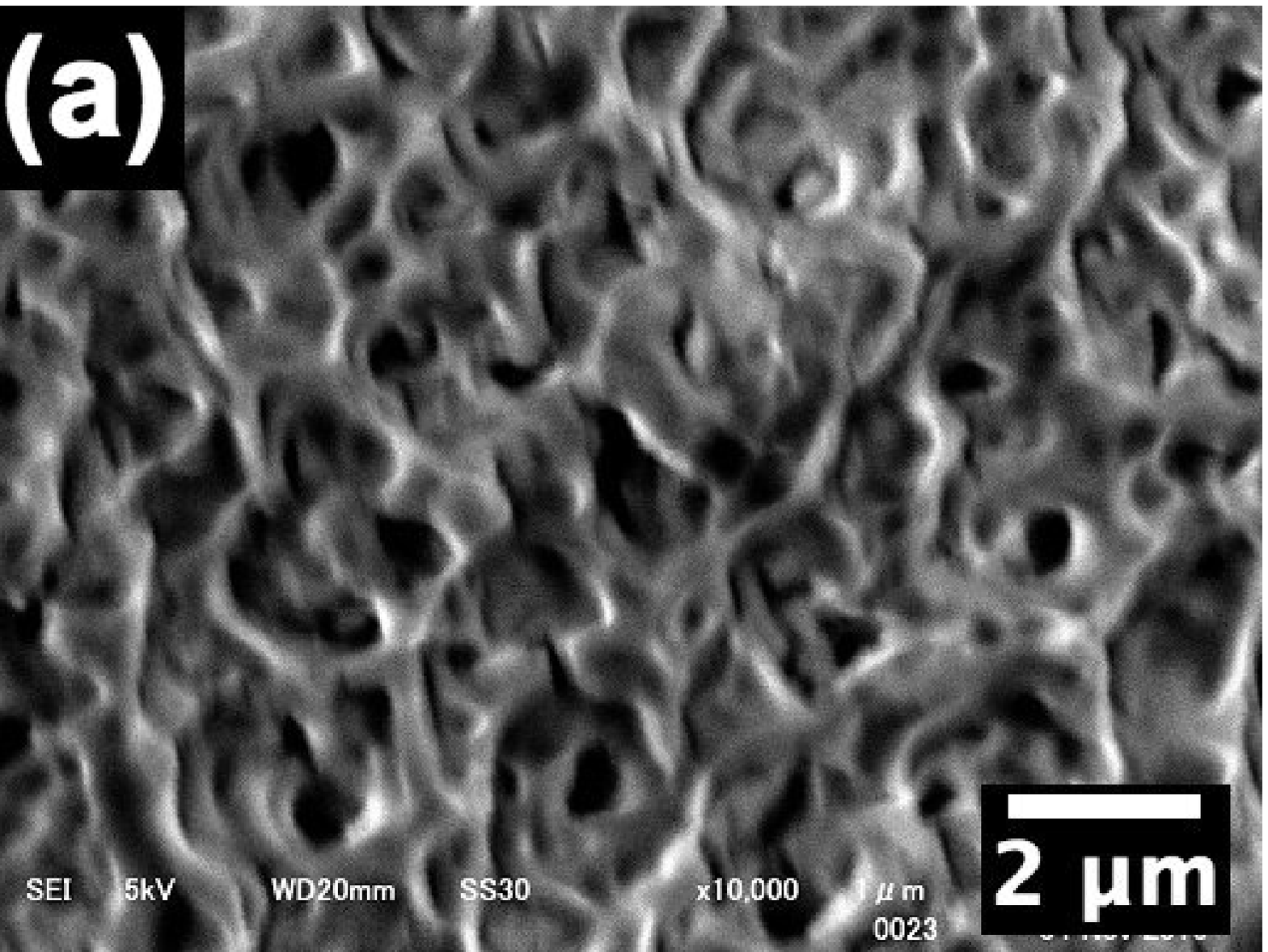}
 \end{flushleft}
 \end{minipage}
 \begin{minipage}{0.33\textwidth}
\begin{flushleft}
 \includegraphics[width=36mm]{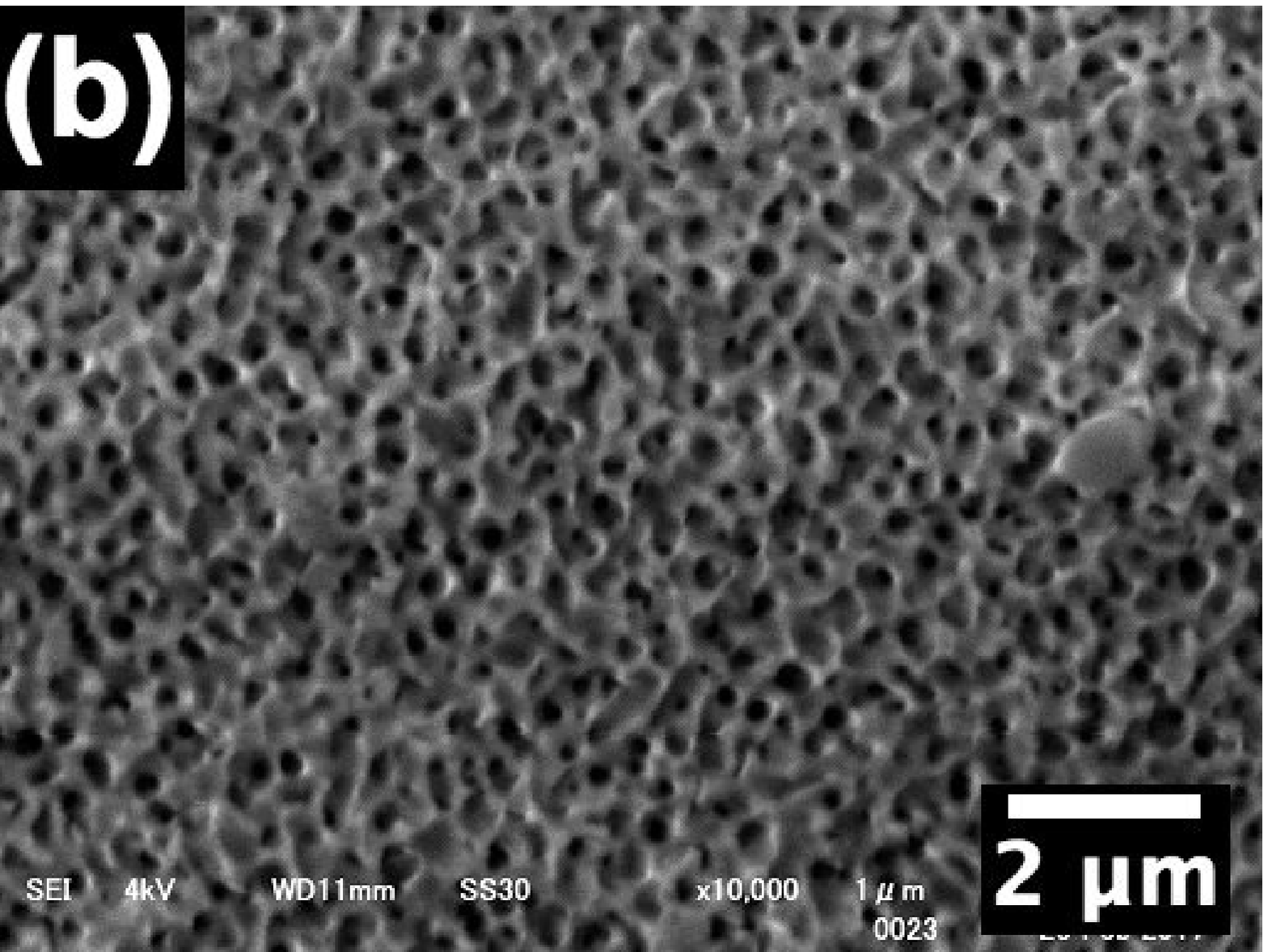}
 \end{flushleft}
 \end{minipage}
 \begin{minipage}{0.3\textwidth}
  \begin{flushleft}
 \begin{enumerate}
 \item[]
  \caption{SEM images of top surfaces of (a) PG-1 and (b) PG-2 containers of Type-2.}
  \end{enumerate}
  \end{flushleft}
 \end{minipage}
  \end{tabular}
 \label{photo-PG}
 \end{figure}

The Type-2 containers are made of phase separated two components of acid-soluble and -insoluble glasses. 
The dimensions are 16/22~mm in inner/outer diameter and 12~mm in inner height. 
The outer surfaces of 36 or 50~$\mathrm{\mu}$m thick were made porous by dissolving the acid-soluble component~\cite{Kukizaki2010}. 
We used two kinds of containers, PG-1 and PG-2, of this type with different pore sizes.
From SEM observations shown in Figs.~\ref{photo-PG}(a),(b), the pore and channel sizes are estimated as 0.5 and 0.3~$\mathrm{\mu}$m for PG-1 and 1.0 and 0.5~$\mathrm{\mu}$m for PG-2, respectively.
Structures of the top surface and cross section of the porous layer are similar to each other.

Surface areas ($S$) of glass pieces prepared in the same way as the Type-1 and -2 containers were determined from  the isothermal adsorption pressure measurements of nitrogen molecules at $T=77$~K. 
All the specifications of the containers are summarized in Table~\ref{table}.

The film flow rate $j$ was determined by measuring the time $t$ during which superfluid $^4$He of a volume $V$ flows out of each container after lifting it up from the superfluid $^4$He bath in a glass dewar. 
$j$ is given by

 \begin{equation}
  j = \frac{V\rho}{tL_{\mathrm{eff}}},
 \label{eqj1}
 \end{equation}

\noindent
where $\rho$ is the density of $^4$He.
We denote $j_0$ rather than $j$ when we assume $L_{\mathrm{eff}} = L_0$ in Eq.~\ref{eqj1}, where $L_0$ is the nominal perimeter of container determined from its macroscopic shape.
For the bare glass container, we can expect $j \approx j_0$.

Since it is not easy to determine the temperature of thin $^4$He film itself, various precautions were taken.
For example, we kept the container bottom close to the bath level (10~mm above) during the flow measurements, and carefully radiation shielded the container region.
Then we assumed that the temperature determined from the $^4$He vapor pressure and a calibrated Allen Bradley resistance thermometer (nominal 200~$\mathrm{\Omega}$) immersed in the bath is reasonably close to the film temperature.
Most of the data were taken at a fixed temperature of $T$=2.04(2)~K. 
The $T$-dependence of $j$ was also measured for PG-1 near $T_\mathrm{\lambda}$ ($1.39 \leq T \leq 2.12$~K).

\section{Results and discussion}

As indicated in Table~1, we observed anomalously large flow rates for Type-1 and -2 containers at $T = 2.04(2)$~K.
They are larger than that for bare glass~\cite{Mendelssohn1950,Smith1955a} by more than two orders of magnitude, and liquid helium flows continuously like a waterfall from the bottom of the container.

\begin{figure}[t]
\begin{minipage}{0.47\textwidth}
\centering
\mbox{\raisebox{10mm}{\includegraphics[width=50mm]{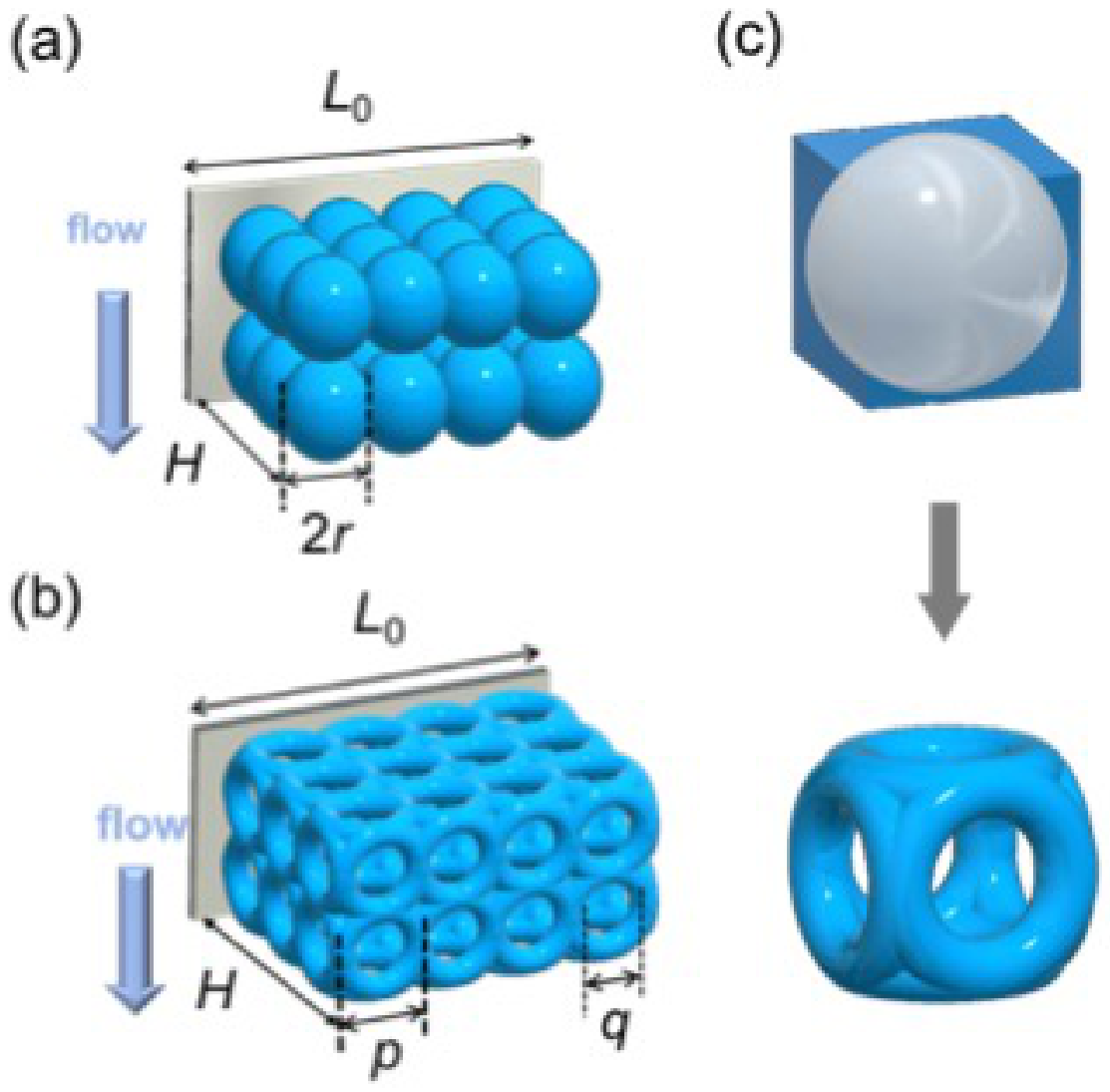}}}
\caption{Schematic drawings of surface structures of (a) Ag powder (Model-1), (b) porous glass (Model-2) and (c) a unit cell in Model-2.}
\label{morphology}
\end{minipage}
\hspace{5mm}
\begin{minipage}{0.47\textwidth}
\centering
\includegraphics[width=50mm]{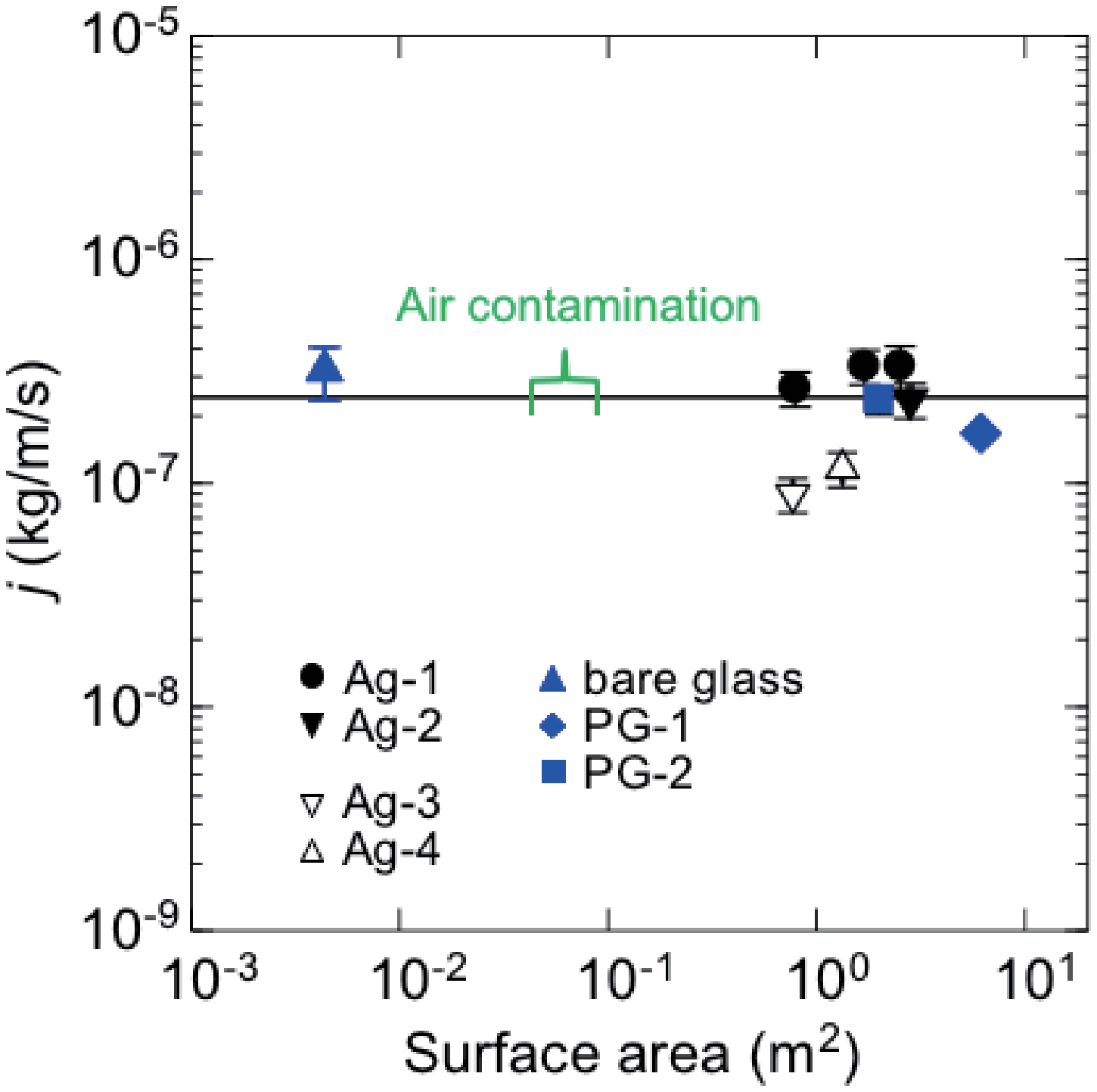}
\caption{
Flow rate $j$ at $T$=2.04~K versus $S$. 
$j$ is obtained through eq.(\ref{eqj1}) by estimating $L_{\mathrm{eff}}$ as described in the main text. 
The horizontal line shows a theoretical value given by eqs.(\ref{eqj2}) and (\ref{eqvc}) with $b=6$~nm, $c=$0.2. 
If the effective perimeter mechanism plays dominant roles in the previous experiments with air contaminated containers~\cite{Bowers1950}, they would be located in the region indicated by the green bracket.
}
\label{j_S}
\end{minipage}
\end{figure}

We estimated the effective perimeter $L_\mathrm{eff}$ by modeling structures of the sintered Ag powder and the porous layers based on the SEM images (Figs.~\ref{photo-Ag} and \ref{photo-PG}). 
For Type-1, the model is packed spheres of $2r = 100$~nm in diameter (see Fig.~\ref{morphology}(a)).
In this model, the total surface area $S$ is given by $4\pi r^2 [S_0 / (2r)^2] [H/2r]$ and $L_\mathrm{eff}$ is given by

\begin{equation}
L_\mathrm{eff}=2\pi r\frac{H}{2r}\frac{L_0}{2r}=\frac{S}{S_0}L_0 \ \ \ \ \ \ \ \ \mathrm{(Model\mathchar`-1)},
\label{eqLeff2}
\end{equation}

\noindent
where $H$ is the thickness of this structure and $S_0$ is the surface area of a base glass container without the Ag powder.
For Type-2, the model is a lattice composed of ring tori (see Fig.~\ref{morphology}(b)).
A unit cell of the lattice consists of 6 tori forming a cube shape.
This is a complementary space left after removing a central sphere from a cube (Fig.~\ref{morphology}(c)).
This jungle-gym structure is likely expected from the fact that the spherically segregated acid-solvable glass component is removed by the dissolving process.
In this model, the surface area of a torus is $\pi^2 [(p-q)/2][(p+q)/2]$ and thus $S/S_{0}$ is given by $(3\pi^2/4)[(p-q)(p+q)H/p^3]$, where $p$ and $q$ are outer and inner diameters of the torus.
Then, $L_\mathrm{eff}$ is given by

\begin{equation}
L_\mathrm{eff}=\pi\frac{p-q}{2}\frac{L_0H}{p^2}=\frac{2}{3\pi} \frac{S}{S_0} \frac{p}{p+q} L_0 \ \ \ \ \ \ \ \ \mathrm{(Model\mathchar`-2)}.
\label{eqLeff3}
\end{equation}
\noindent
$L_\mathrm{eff}$ values estimated by these models are tabulated in Table~\ref{table}.

Using these $L_\mathrm{eff}$ values, we can deduce $j$ for various containers from Eq.\ref{eqj1}.
They are plotted in Fig.~\ref{j_S} with respect to $S$.
All the $j$ data, except for Ag-3 and 4, fall on the same value within a factor of two. 
This is a surprisingly good agreement considering such different surface morphology and surface areas over two orders of magnitude.
Clearly, the enhancement of $v_{\mathrm{c}}$ has a negligibly small effect on the dramatic increase of $j_0$, and $v_{\mathrm{c}}$ changes only a little, within a factor of 3--4 at most, regardless of such largely different surface roughness.
In other words, the geometrical effect plays a dominant role on the observed enhancement of $\rho_0$ in our surface decorated containers, and this must be the most relevant mechanism in the containers contaminated with solid air.

The data for Ag-3 and 4 seem to be systematically smaller than the others approximately by a factor of two (see Fig.~\ref{j_S}).
This is presumably due to poorer applicability of the Model-2 to the case when insufficient neck grows between adjacent Ag particles due to the lower $T_{\mathrm{sint}}$ and the shorter $t_{\mathrm{sint}}$.
Such thin neck parts will contribute to 
overestimate $L_\mathrm{eff}$.
The He film thickness adsorbed on Ag surfaces should be larger than that on glass surfaces by about 40\% because of the stronger Van der Waals attractive force.
However, such a small difference is not important in this study where we varied $L_\mathrm{eff}$ to so large extent.

The film flow rates are nearly independent of the liquid level in our experiment as in the previous works~\cite{Daunt1954,Duthler1971}.
From this fact it is natural to consider that the flows are also determined by the superfluid critical velocity $v_{\mathrm{c}}$.
If we assume the constant film thickness $\eta$ over the whole surface, which is about 30~nm on glass surfaces and is weakly level dependent eventually, $V/t \approx v_{\mathrm{c}} \eta L_{\mathrm{eff}} \rho_{\mathrm{s}}/\rho$. 
Here $\rho_{\mathrm{s}}/\rho$ is the superfluid fraction which is well tabulated as a function of $T$ in literature~\cite{Donnelly2009}.
Then Eq.~\ref{eqj1} can be rewritten as
 
\begin{equation}
  j (T) = v_{\mathrm{c}} \eta \rho_\mathrm{s} (T).
 \label{eqj2}
 \end{equation} 
 
\noindent
and through this relation $v_{\mathrm{c}}$ can be deduced from $j$.

In the early days, it has been considered that $v_{\mathrm{c}}$ in the film flow is $T$-independent
~\cite{Daunt1954}.
However, as far as we know, it has not been reexamined carefully from the view point of subsequent theoretical progress.
Among others, the vortex depinning model by Schwarz~\cite{Schwarz1985} will be one of the most relevant descriptions for $v_{\mathrm{c}}$ in our experimental setup.
In this model, $v_{\mathrm{c}}$ is given by

 \begin{equation}
   v_{\mathrm{c}} (T) = c \frac{h}{4\pi m \eta} \ln {(b/a (T))},
 \label{eqvc}
 \end{equation}

\noindent
where $h$, $m$, $b$, and $a$ are Planck's constant, mass of He atom, typical size of surface pinning sites, and the vortex core radius ($\sim$0.1~nm).
$c$ is a constant of the order of unity.
The insensitivity of $v_{\mathrm{c}}$ to the surface decoration we observed can be explained by the logarithmically weak dependence of $v_{\mathrm{c}}$ on $b$ within this model.
Since $a$ has weak $T$-dependence except for close to $T_\mathrm{\lambda}$~\cite{Barenghi1983}, the relevancy of the vortex depinning model can be tested if we deduce $v_{\mathrm{c}} (T)$ from $j$ measured at various $T$ near $T_\mathrm{\lambda}$ through Eq.~\ref{eqj2} and compare it with Eq.~\ref{eqvc}.

\begin{figure}[b]
\begin{minipage}{0.4\textwidth}
\centering
\includegraphics[width=50mm]{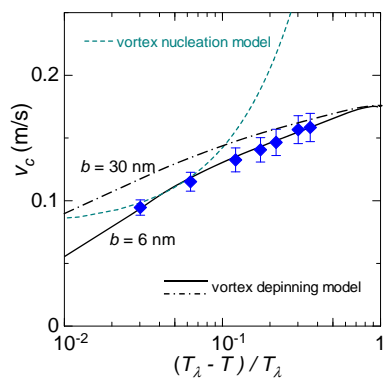}
\end{minipage}
\hspace{10mm}
\begin{minipage}{0.4\textwidth}
\begin{enumerate}
\item[] \caption{$T$-dependence of $v_{\mathrm{c}}$ deduced from the film flow experiment for Type-2 container. Fittings to Eq.~\ref{eqvc} with $b = 6$~nm, $c = 0.16$ and $b = 30$~nm, $c = 0.12$ are indicated by the solid line and the dash-dot one, respectively. The dashed line is a fitting to the vortex nucleation model~\cite{Varoquaux1986, Beecken1987}. See text for details.}
\end{enumerate}
\end{minipage}
\label{figjT}
\end{figure}%
\setlength\textfloatsep{2truemm}

Eventually,  we found a weak but clear $T$-dependence of $v_{\mathrm{c}}$ for PG-1 container in the whole $T$ region ($1.39 \leq T \leq 2.12$~K ) we studied (see Fig.~5).
Moreover, the $T$-dependence of $v_{\mathrm{c}}$ data can be well represented by Eq.~\ref{eqvc} with $b = 6$~nm and $c = 0.16$ as indicated by the solid line where we used the $T$-dependence of $a$ given in Ref.~\cite{Barenghi1983}.
On the other hand, a fitting to Eq.~\ref{eqvc} with $b = 30$~nm (the long dashed dotted line; $c = 0.12$) does not represent the data well.
Also, a fitting to another formula $v_{\mathrm{c}} = c\{1-(T/T_0)\}$, which is proposed by the vortex nucleation mechanism~\cite{Varoquaux1986, Beecken1987}, (the dashed line; $c = 0.22$, $T_0 = 2.45$~K) can not represent the data at all.
Thus our results strongly suggest that the superfluid film flow is determined by depinning of remnant vortices from pinning sites smaller than 10~nm.
Further experiments extending $T$ range with improved precisions and most expectedly for containers made of other materials will be valuable to confirm this conclusion.

Finally, we comment that similarly deduced $v_{\mathrm{c}}(T)$ from $j$ data in the previous film flow experiments scatter too largely to draw a clear conclusion. 
Some of them seem to show even a stronger $T$-dependence~\cite{Daunt1939} or almost no dependence~\cite{Mendelssohn1950}.

\section{Conclusions}

We have measured the apparent film flow rate $j_0$ of superfluid $^4$He for glass containers with different surface decorations. 
With increasing the surface area, i.e., the effective perimeter $L_\mathrm{eff}$, we observed a huge enhancement of $j_0$ by more than two orders of magnitude. 
By properly evaluating $L_\mathrm{eff}$ based on the simple models for the surface morphology, we concluded that the enhancement is almost throughly caused by the geometrical effect not by the substantial increase of the critical velocity $v_{\mathrm{c}}$. Larger flow rate of the air contaminated container seems to be explained by the increase of the effective perimeter. 
From the measurement of temperature dependence of the flow rate, it was strongly suggested that $v_{\mathrm{c}}$ is weakly $T$ dependent in accordance with the vortex depinning mechanism.

\begin{acknowledgements}

 \ The authors thank Akagawa Glass Co. Ltd. for providing us the porous glass containers. We are also grateful to the late Tadao Imai for processing the Pyrex glass containers. JU was supported by Japan Society for the Promotion of Science through Program for Leading Graduate Schools (MERIT).

\end{acknowledgements}

\bibliographystyle{spphys}       
\bibliography{Ref.bib}   

\begin{thebibliography}{10}
\providecommand{\url}[1]{{#1}}
\providecommand{\urlprefix}{URL }
\expandafter\ifx\csname urlstyle\endcsname\relax
  \providecommand{\doi}[1]{DOI \discretionary{}{}{}#1}\else
  \providecommand{\doi}{DOI \discretionary{}{}{}\begingroup
  \urlstyle{rm}\Url}\fi

\bibitem{Daunt1938}
J.G. Daunt, K.~Mendelssohn, Nature \textbf{141}(3577), 911 (1938).
\newblock \doi{10.1038/141911a0}.
\newblock \urlprefix\url{http://www.nature.com/articles/141911a0}

\bibitem{Everitt1964}
C.W.F. Everitt, K.R. Atkins, A.~Denenstein, Physical Review \textbf{136}(6A),
  A1494 (1964).
\newblock \doi{10.1103/PhysRev.136.A1494}.
\newblock \urlprefix\url{https://link.aps.org/doi/10.1103/PhysRev.136.A1494}

\bibitem{Bowers1950}
R.~Bowers, K.~Mendelssohn, Proceedings of the Physical Society. Section A
  \textbf{63}(12), 1318 (1950).
\newblock \doi{10.1088/0370-1298/63/12/303}.
\newblock
  \urlprefix\url{http://stacks.iop.org/0370-1298/63/i=12/a=303?key=crossref.be962ad581b38b2a211580a85fddbd9a}

\bibitem{Chandrasekhar1952}
B.S. Chandrasekhar, K.~Mendelssohn, Proceedings of the Physical Society.
  Section A \textbf{65}(3), 226 (1952).
\newblock \doi{10.1088/0370-1298/65/3/111}.
\newblock
  \urlprefix\url{http://stacks.iop.org/0370-1298/65/i=3/a=111?key=crossref.5a45b692ae11f9c3c68c27c1b32939d5}

\bibitem{Smith1955}
B.~Smith, H.A. Boorse, Physical Review \textbf{99}(2), 346 (1955).
\newblock \doi{10.1103/PhysRev.99.346}.
\newblock \urlprefix\url{https://link.aps.org/doi/10.1103/PhysRev.99.346}

\bibitem{Mendelssohn1950}
K.~Mendelssohn, G.K. White, Proceedings of the Physical Society. Section A
  \textbf{63}(12), 1328 (1950).
\newblock \doi{10.1088/0370-1298/63/12/304}.
\newblock
  \urlprefix\url{http://stacks.iop.org/0370-1298/63/i=12/a=304?key=crossref.f8c9ed9650a563dc731f4c60d2a35e4e}

\bibitem{Smith1955a}
B.~Smith, H.A. Boorse, Physical Review \textbf{98}(2), 328 (1955).
\newblock \doi{10.1103/PhysRev.98.328}.
\newblock \urlprefix\url{https://link.aps.org/doi/10.1103/PhysRev.98.328}

\bibitem{Schwarz1985}
K.W. Schwarz, Physical Review B \textbf{31}(9), 5782 (1985).
\newblock \doi{10.1103/PhysRevB.31.5782}.
\newblock \urlprefix\url{https://link.aps.org/doi/10.1103/PhysRevB.31.5782}

\bibitem{Kukizaki2010}
M.~Kukizaki, Journal of Membrane Science \textbf{360}(1-2), 426 (2010).
\newblock \doi{10.1016/J.MEMSCI.2010.05.042}.
\newblock
  \urlprefix\url{https://www.sciencedirect.com/science/article/pii/S0376738810004102}

\bibitem{Daunt1954}
J.G. Daunt, R.S. Smith, Reviews of Modern Physics \textbf{26}(2), 172 (1954).
\newblock \doi{10.1103/RevModPhys.26.172}.
\newblock \urlprefix\url{https://link.aps.org/doi/10.1103/RevModPhys.26.172}

\bibitem{Duthler1971}
C.J. Duthler, G.L. Pollack, Physical Review A \textbf{3}(1), 191 (1971).
\newblock \doi{10.1103/PhysRevA.3.191}.
\newblock \urlprefix\url{https://link.aps.org/doi/10.1103/PhysRevA.3.191}

\bibitem{Donnelly2009}
R.J. Donnelly, C.F. Barenghi, Journal of Physical and Chemical Reference Data
  \textbf{27}(6), 1217 (1998).
\newblock \doi{10.1063/1.556028}.
\newblock \urlprefix\url{http://aip.scitation.org/doi/abs/10.1063/1.556028
  http://scitation.aip.org/content/aip/journal/jpcrd/27/6/10.1063/1.556028}

\bibitem{Barenghi1983}
C.F. Barenghi, R.J. Donnelly, W.F. Vinen, Journal of Low Temperature Physics
  \textbf{52}(3-4), 189 (1983).
\newblock \doi{10.1007/BF00682247}.
\newblock \urlprefix\url{http://link.springer.com/10.1007/BF00682247}

\bibitem{Varoquaux1986}
E.~Varoquaux, M.W. Meisel, O.~Avenel, Physical Review Letters \textbf{57}(18),
  2291 (1986).
\newblock \doi{10.1103/PhysRevLett.57.2291}.
\newblock \urlprefix\url{https://link.aps.org/doi/10.1103/PhysRevLett.57.2291}

\bibitem{Beecken1987}
B.P. Beecken, W.~Zimmermann, Physical Review B \textbf{35}(4), 1630 (1987).
\newblock \doi{10.1103/PhysRevB.35.1630}.
\newblock \urlprefix\url{https://link.aps.org/doi/10.1103/PhysRevB.35.1630}

\bibitem{Daunt1939}
J.G. Daunt, K.~Mendelssohn, Proceedings of the Royal Society of London. Series
  A, Mathematical and Physical Sciences \textbf{170}(942), 439 (1939).
\newblock \urlprefix\url{http://www.jstor.org/stable/97282}

\end{thebibliography}

%
%

\end{document}